\documentclass{vgtc}                          

\ifpdf
  \pdfoutput=1\relax                   
  \usepackage{graphicx}                
  \DeclareGraphicsExtensions{.pdf,.png,.jpg,.jpeg} 
\else
  \usepackage{graphicx}                
  \DeclareGraphicsExtensions{.eps}     
\fi%

\graphicspath{{figures/}{pictures/}{images/}{./}} 

\usepackage{microtype}                 
\PassOptionsToPackage{warn}{textcomp}  
\usepackage{textcomp}                  
\usepackage{mathptmx}                  
\usepackage{times}                     
\usepackage{cite}                      
\usepackage{tabularx}                  
\usepackage{booktabs}                  
\usepackage{subcaption}
\usepackage{comment}
\usepackage{amsmath}

\onlineid{0}

\title{Climate Physicalization: Bridging Tangible Interaction and Data Visualization for Environmental Communication}

\author{Milena Vuckovic\thanks{e-mail: vuckovic@vrvis.at}\\ %
        \scriptsize VRVis GmbH %
\and Marlene Huber\thanks{e-mail: mhuber@vrvis.at}\\ %
     \scriptsize VRVis GmbH \\ \scriptsize TU Wien %
\and Fabrizia Bando-Bechtold\thanks{e-mail: bechtold@vrvis.at}\\ %
     \scriptsize VRVis GmbH %
\and Daniel Cornel\thanks{e-mail: cornel@vrvis.at}\\ %
     \scriptsize VRVis GmbH %
\and Daniel Pahr\thanks{e-mail: daniel.pahr@univie.ac.at}\\ %
     \scriptsize University of Vienna} %

\abstract{
Raising awareness of climate topics, such as urban heat mitigation, requires communicating complex data to a broad audience. Data physicalization has emerged as a novel method to connect audiences to such complex topics through tangible forms while also leveraging interactivity to foster deeper engagement.
This paper introduces an situated geospatial thermal physicalization prototype, designed to enhance communication and education of environmental information through hands-on engagement. The physical model depicts a 3D model of an urban neighborhood, with radiation data represented as a layered contour structure. The model incorporates three interchangeable climate adaptation scenarios, alongside an externally mounted, electronically driven solar path representation featuring multiple discrete LEDs that can be switched on and off at will. To investigate how people engage with the model, we conducted a hands-on qualitative lab study with 16 participants, using think-aloud, observational, and questionnaire protocols. Overall, participants responded very favorably to the model, highlighting the novelty of the visualization approach and its capacity to improve the interpretability of the presented information.
} 

\begin{document}


\firstsection{Introduction}

\maketitle

Climate visualization is a critical medium for effective environmental communication in both scientific and public discourse~\cite{johannson2010}. By transforming complex, multidimensional climate data into interpretable visual formats, such as geospatial maps, 2D heatmaps, and graphs, we can bridge the gap between raw scientific findings and actionable insights through visualization~\cite{Terrado2022,Monoyios2024}. While contemporary climate visualization increasingly adopts interactive, exploratory 3D formats, the field continues to rely on static 2D representations. This reliance on 2D formats may obscure the nuanced patterns in the data, as the flattening of dynamic, spatial information may result in substantial loss of detail~\cite{Schneider2012,Rober2021}. Consequently, these abstractions not only oversimplify complex datasets but also trivialize the gravity of climate change~\cite{Schuster2024}. This limitation is particularly critical for public communication, as audiences without specialized knowledge may struggle to accurately interpret even the simplest visualization formats, let alone the underlying data~\cite{vuckovic2025}. Given the urgent need for collective action to address the growing challenges of climate change, it is equally critical that everyone not only has access to climate information but is also empowered to act upon it.

In response to this, different innovative approaches to environmental communication are gaining momentum, with data physicalization emerging as a particularly promising method. Data physicalization focuses on transforming abstract digital data into its tangible, physical counterpart, enabling people to explore, understand, and interact with information in novel, multisensory, and intuitive ways~\cite{Jansen2015}. Essentially, these physicalizations are meant to externalize abstract concepts, allowing us to engage with, sense, and emotionally connect to data through real objects that embody the respective information~\cite{Sauve2023}. The potential applications of these data representations are gradually being explored in the field of environmental communication~\cite{Wang2019,Perovich2021a,Pittarello2024}. Yet, the majority of these applications still focus solely on abstract, non-situated representations of environmental information, rather than embedding the data within its natural, real-world context. Additionally, many of the physical artifacts discussed in the literature are inherently static, relying exclusively on their geometric shapes or material properties to communicate data. This static nature limits user engagement and sensemaking~\cite{Pahr2024}, offering only a fixed perspective on the information it represents. 

To tackle these limitations, our work introduces \textbf{a novel approach to climate physicalization}. By embedding geospatial thermal data into an architectural-scale model, we created a situated geospatial thermal physicalization prototype featuring a layered contour structure where distinct temperature ranges are discretely encoded into the height of each contour layer. This design not only visualizes data but also invites haptic exploration and deeper engagement with the information, specifically, heat distribution within an urban neighborhood and selected strategies to mitigate its negative effects. Interactivity is achieved by integrating selected climate adaptation strategies into the model, allowing users to explore and exchange different scenarios. Additionally, the physical model provides a tangible metaphor of a key driver of radiation exchange -- individual sun positions along the solar path. Users can interactively manipulate these sun positions by switching them on and off and dynamically explore their influence on microclimate development, thereby gaining an intuitive, hands-on understanding.

We may summarize the primary contributions of our work as follows:
\begin{itemize}
\item \textbf{Fabrication process and design decisions} for a situated physical urban climate scale model with embedded electronics;
\item \textbf{Strategies for introducing interactivity} in a physical climate data model through direct user manipulation of physical artifacts and electronics;  
\item \textbf{Insights on user engagement} derived from an empirical investigation of interactions with the physical data model.
\end{itemize}

\section{Related Work}

\paragraph{Visualizing Climate Data} In the context of climate data visualization for practitioners, the focus is primarily on representing complex datasets and underlying physical processes across global, regional, and local scales. At the global and regional levels, this typically involves depicting extensive spatial domains, often mapped onto a globe or defined geographic region~\cite{Schneider2012,Rober2021}. Such representations may employ various projection methods and are frequently limited to a select set of variables, for instance, atmospheric dynamics, risk and hazard indicators, or measures of uncertainty. At the local scale, visualizations may encompass either the entire urban area or a specific neighborhood~\cite{Huber2025}. In such cases, they typically concentrate on a single variable at a time, such as air temperature, wind velocities, solar exposure, or thermal comfort indices~\cite{Rober2013}. However, much of public-facing climate communication continues to rely on the representation of linear trends through line charts, rather than depicting complex spatial dynamics and patterns of change, which are more commonly employed in practitioner-oriented analyses~\cite{Schuster2024}. While such visualizations are effective for conveying long-term changes and general trajectories, they often reduce complex, multidimensional processes to simplified patterns. Additionally, these approaches predominantly rely on static representations, while the adoption of interactive visualization methods remains comparatively limited due to the inherent complexity of climate phenomena~\cite{Tominski2011}.

\paragraph{Environmental Data Physicalization} As mentioned at the outset, the application of data physicalization in the environmental domain is still in its infancy. Only a limited number of studies have explored such physical representations, with most relying primarily on anchoring data to non-situated geometric forms~\cite{Wang2019,Perovich2021a,Perovich2021b,Pittarello2024}. 

To exemplify, Pittarello and Semenzato~\cite{Pittarello2024} described a study in which participants constructed a physical representation of environmental data gathered from real-world neighborhoods. The result, however, was a contextually detached cylindrical model that mapped the collected data along an abstracted version of the path the participants traversed. Perovich et al.~\cite{Perovich2021b} explored the use of human-sized bar charts to support collective discussion and engagement with indoor environmental exposure data, particularly household pollutant levels. Similar to the previous study, this installation was conceived as a non-situated sculptural artifact. Yet, in their other study, Perovich et al.~\cite{Perovich2021a} experimented with the contextually anchored, situated data physicalization by visualizing chemical water pollution directly on the affected water body. However, even in this case, the medium used to represent the data relied on highly abstracted forms, rather than on an artifact that externalizes the very nature of the data being communicated. 

\paragraph{Situated Data Representations} Situated visualizations refer to a practice of embedding data representations directly within physical spaces, objects, and environments that are contextually pertinent to the data itself~\cite{Bressa2021}. Although often associated with immersive visualization, situated data representations are broadly applicable to a wide range of context-sensitive applications. This is especially pertinent in scenarios where embedding data within its relevant physical or environmental context significantly enhances both its interpretability and practical applicability. Nevertheless, a substantial volume of existing applications continues to be confined to immersive environments~\cite{White2009,Zollmann2016,Martins2022,Nagel2024}, with comparatively limited exploration of applications beyond this domain~\cite{Santos2024}.

\paragraph{Physical Scale Models} 
3D relief maps are common as exhibition pieces in museums and public displays, and projection augmentation is a common way to add dynamic data to such artifacts~\cite{priestnall_projection_2012}. Alternatively, data can be encoded into height~\cite{Rase2012} or texture of maps~\cite{djavaherpour_first_2023} to leverage the third spatial dimension.
However, the use of architectural scale models remains largely confined to architectural practice. As discussed by Hull and Willet~\cite{Hull2017}, these models have the capacity to convey depth, texture, and materiality, yet their application is predominantly limited to supporting building and landscape design processes alone. To the best of our knowledge, there is no comparable study that adds additional layers of simulated environmental information (e.g., geospatial thermal data) on top of a basic structural, architectural model of a real-world domain through physicalization, while also leveraging tangible interaction with interchangeable scenarios.  

\paragraph{Tangible Interaction} Interactive data physicalizations offer significant potential to foster dynamic, embodied experiences in which users can actively explore and manipulate data through real-time, tangible interaction. Previous research has explored various approaches to interactive physicalization, distinguishing between direct and indirect manipulation of elements. In this study, our focus lies specifically on direct user interaction, rather than indirect interaction or the inherent interactive properties of physicalizations themselves. Some examples of such direct interaction relate to squeezing~\cite{Hull2017,Pahr2025a}, pushing~\cite{Taher2017,Pahr2024}, touching~\cite{Bae2024}, or spatially arranging the physical elements~\cite{huron_constructing_2014, Pahr2025b}. However, as discussed by Bae et al.~\cite{Bae2024}, creating interactive physicalizations is inherently challenging because it requires cross-disciplinary knowledge spanning visualization, fabrication, and, in some cases, electronics. The fundamentally transient and dynamic nature of environmental phenomena adds a further level of complexity to their translation into physical representations. This may partly explain why such approaches remain relatively underexplored in the context of environmental applications.

\section{Methodology}

The process of constructing and evaluating a situated, physical geospatial thermal model followed a phased approach that consisted of the following steps: (1) obtaining the underlying 3D urban and simulated thermal data from an environmental numerical model; (2) fabricating individual physical artifacts of the scaled model; (3) conducting an empirical study to investigate user behavior in conjunction with our apparatus.

\subsection{Numerical simulation}

Our research focuses on the practical application of visualizing and interacting with selected climate adaptation strategies and the resulting local climate development in a sample urban neighborhood. We considered three distinct climate adaptation scenarios: (1) baseline conditions without trees and fully exposed to incoming solar radiation; (2) a single-tree configuration to illustrate the effect of radiation shielding by one tree; and (3) a multiple-tree configuration to illustrate the effect of radiation shielding by multiple trees.

\begin{figure*}[tb]
 \centering 
 \begin{subfigure}{0.32\textwidth}
        \centering
        \includegraphics[width=\linewidth]{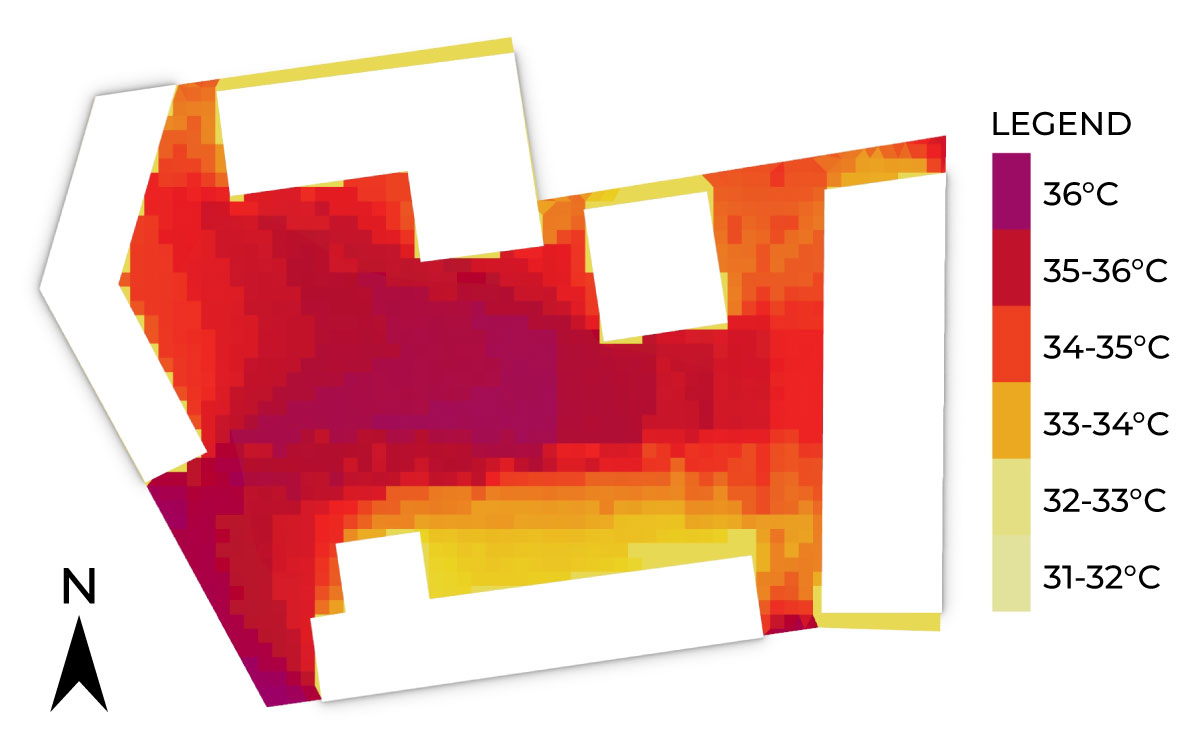}
    \end{subfigure}
    \hfill
\begin{subfigure}{0.32\textwidth}
        \centering
        \includegraphics[width=\linewidth]{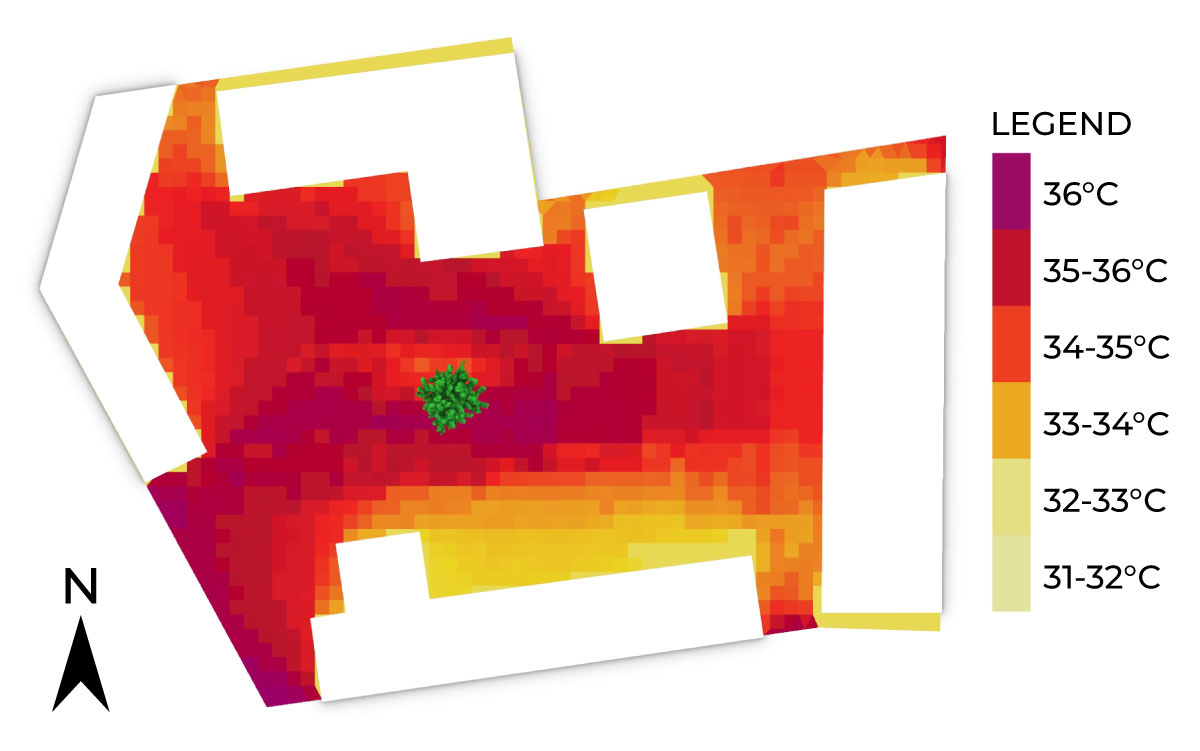}
    \end{subfigure}
    \hfill
\begin{subfigure}{0.32\textwidth}
        \centering
        \includegraphics[width=\linewidth]{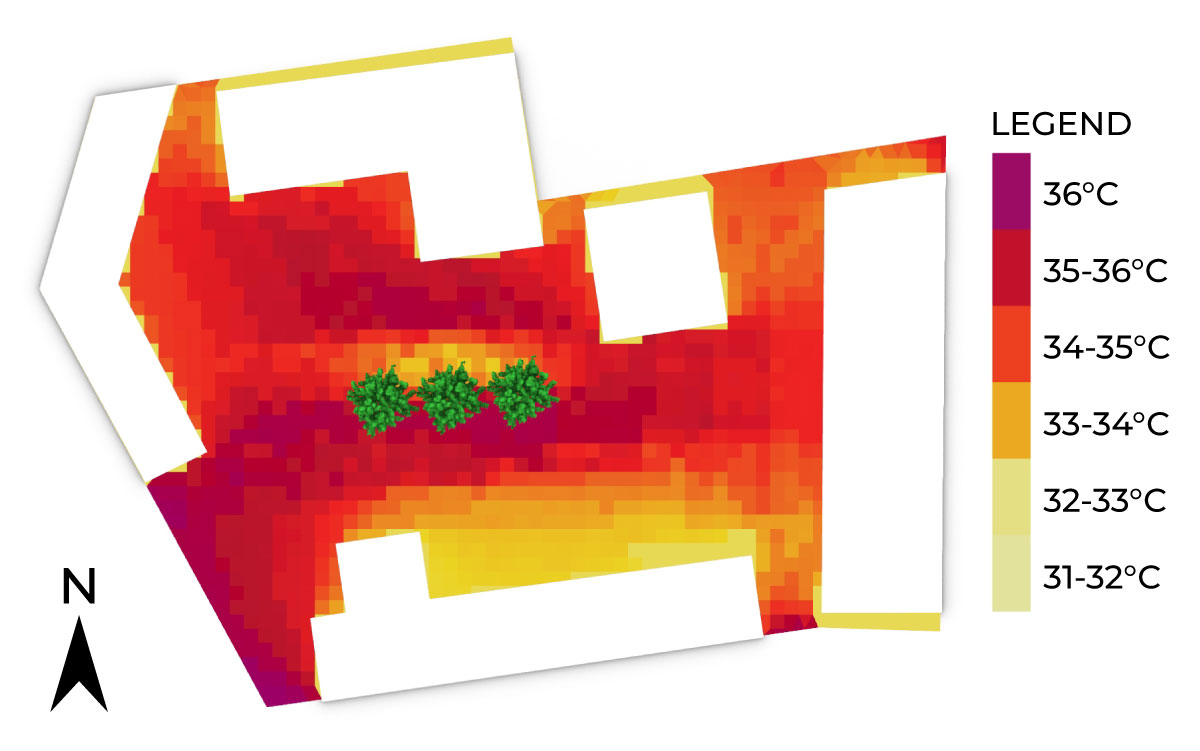}
    \end{subfigure}
 \caption{Overview of three simulated climate adaptation scenarios and the resulting thermal environment: (left) baseline condition without trees; (middle) single-tree configuration; (right) multi-tree arrangement.}
 \label{fig:sims}
\end{figure*}

We employed the Rhinoceros 3D/Grasshopper parametric modeling environment, along with its Ladybug plug-in and the associated tool library, to perform high-resolution thermal simulations for each of the considered scenarios~\cite{TRUBIANO2013,Vuckovic2019}. The tools from the Ladybug library couple the 3D modeling environment with validated energy and daylighting simulation engines. The 3D urban and temporal boundary weather-forcing data were obtained from open-source repositories for an urban area in Vienna, Austria. The simulations were run for a selected hot summer day in August, with the modeling results averaged over the simulated daytime hours (Figure~\ref{fig:sims}). Although the Ladybug library enables the calculation of various thermal comfort indices, we focused on the Universal Thermal Climate Index (UTCI), a widely used standard for assessing outdoor thermal comfort~\cite{Broede2010}. Expressed in degrees Celsius, UTCI is designed to be readily interpretable by non-expert audiences, making it particularly suitable for public-facing applications.

\subsection{Fabrication of a physical model}

The fabrication process for the physical model was guided by design considerations regarding context, structure, and interactions, as set out by Bae et al.~\cite{Bae2022}. The process itself was a collaborative effort led by the principal researchers, who worked closely to identify suitable methods for encoding the data. 

\begin{figure*}[tb]
 \centering 
 \begin{subfigure}{0.32\textwidth}
        \centering
        \includegraphics[width=\linewidth]{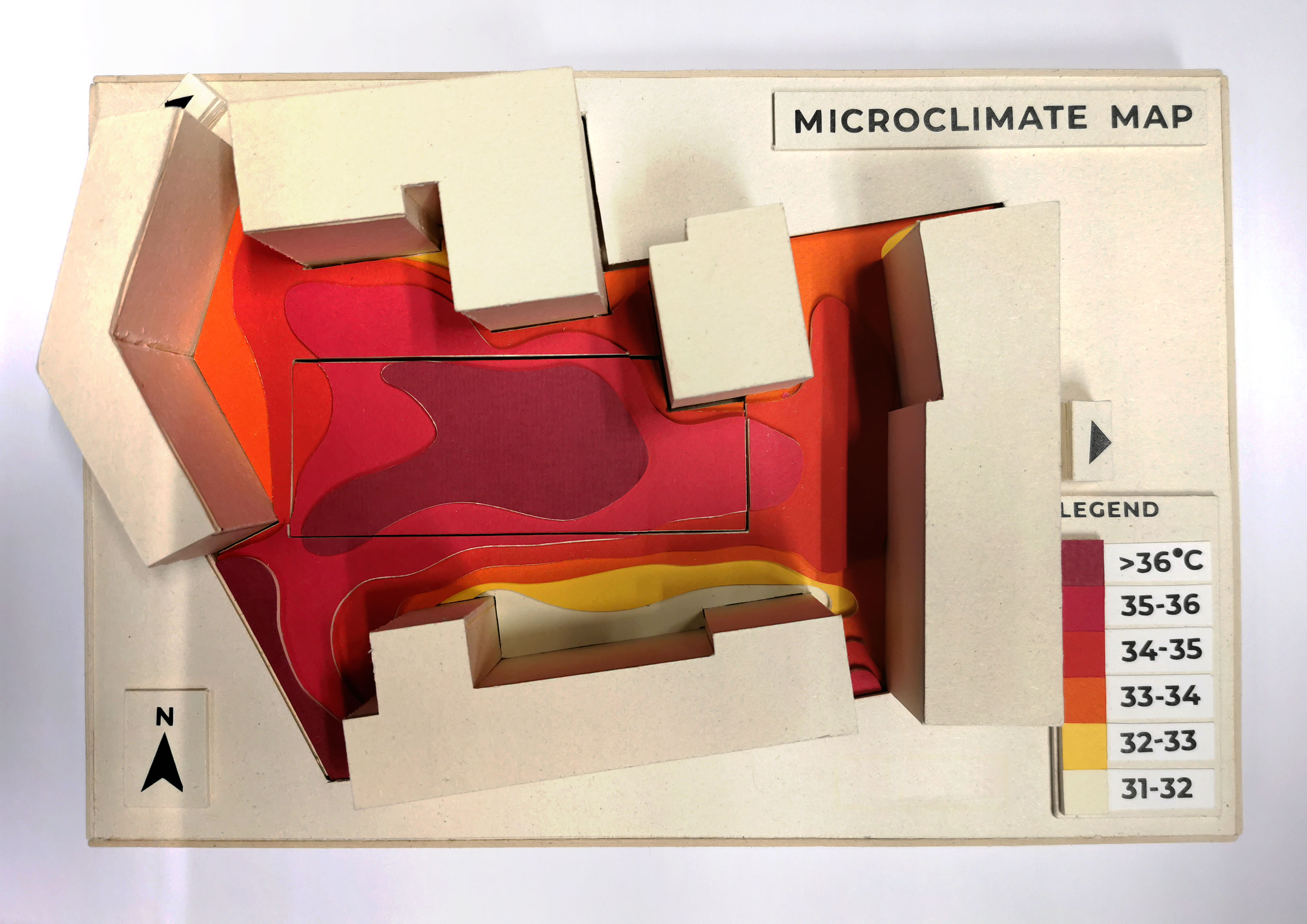}
    \end{subfigure}
    \hfill
\begin{subfigure}{0.32\textwidth}
        \centering
        \includegraphics[width=\linewidth]{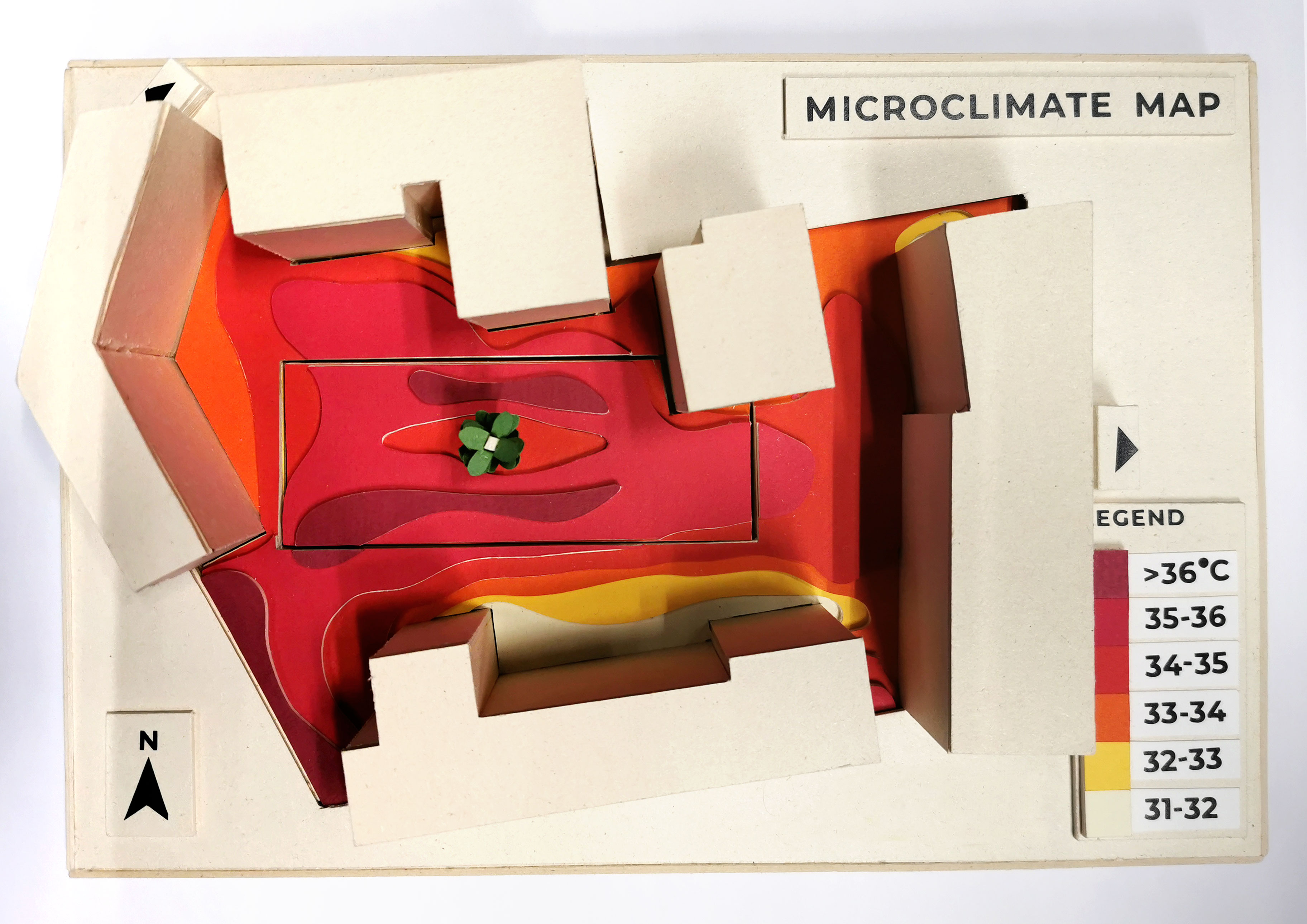}
    \end{subfigure}
    \hfill
\begin{subfigure}{0.32\textwidth}
        \centering
        \includegraphics[width=\linewidth]{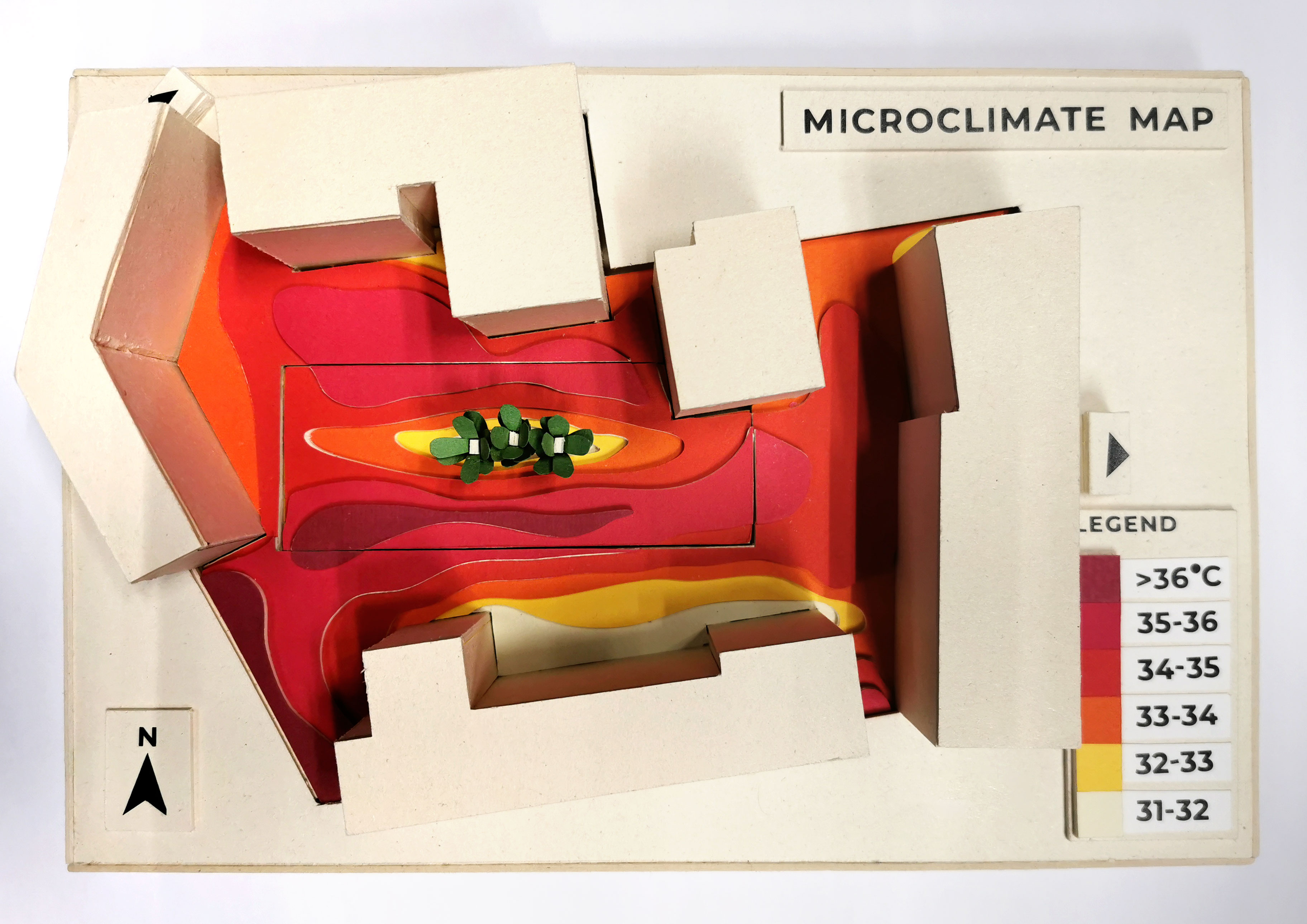}
    \end{subfigure}
 \caption{The physical model of the urban neighborhood with two exchangeable inlays, each representing one of the considered tree configuration scenarios: (left) baseline condition without trees; (middle) single-tree configuration; (right) multiple-tree arrangement.}
 \label{fig:physicalModel}
\end{figure*}

\begin{figure}[tb]
 \centering 
 \begin{subfigure}{0.45\textwidth}
        \centering
        \includegraphics[width=\linewidth]{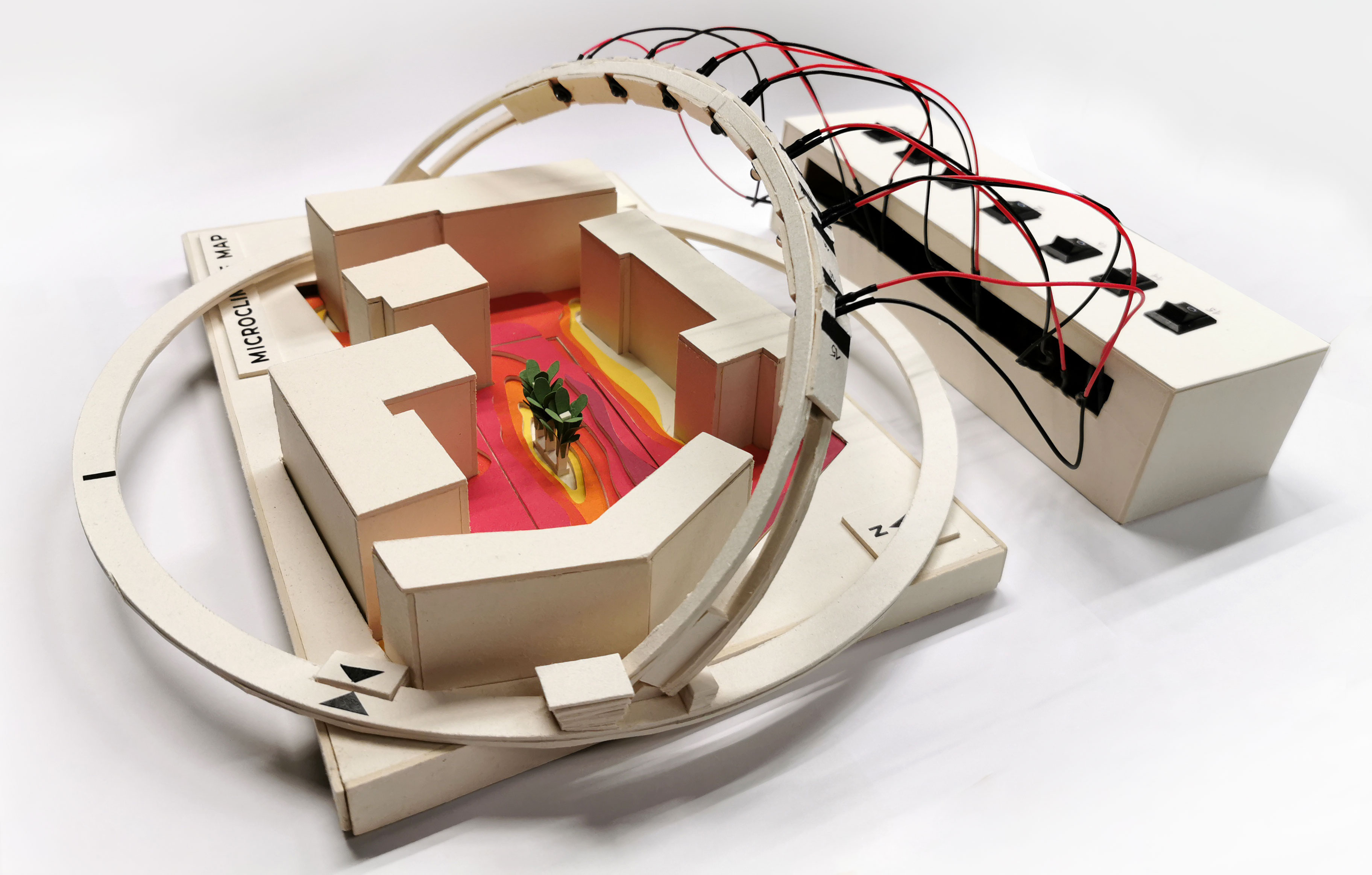}
    \end{subfigure}
    \par\vspace{0.25em} 
\begin{subfigure}{0.45\textwidth}
        \centering
        \includegraphics[width=\linewidth]{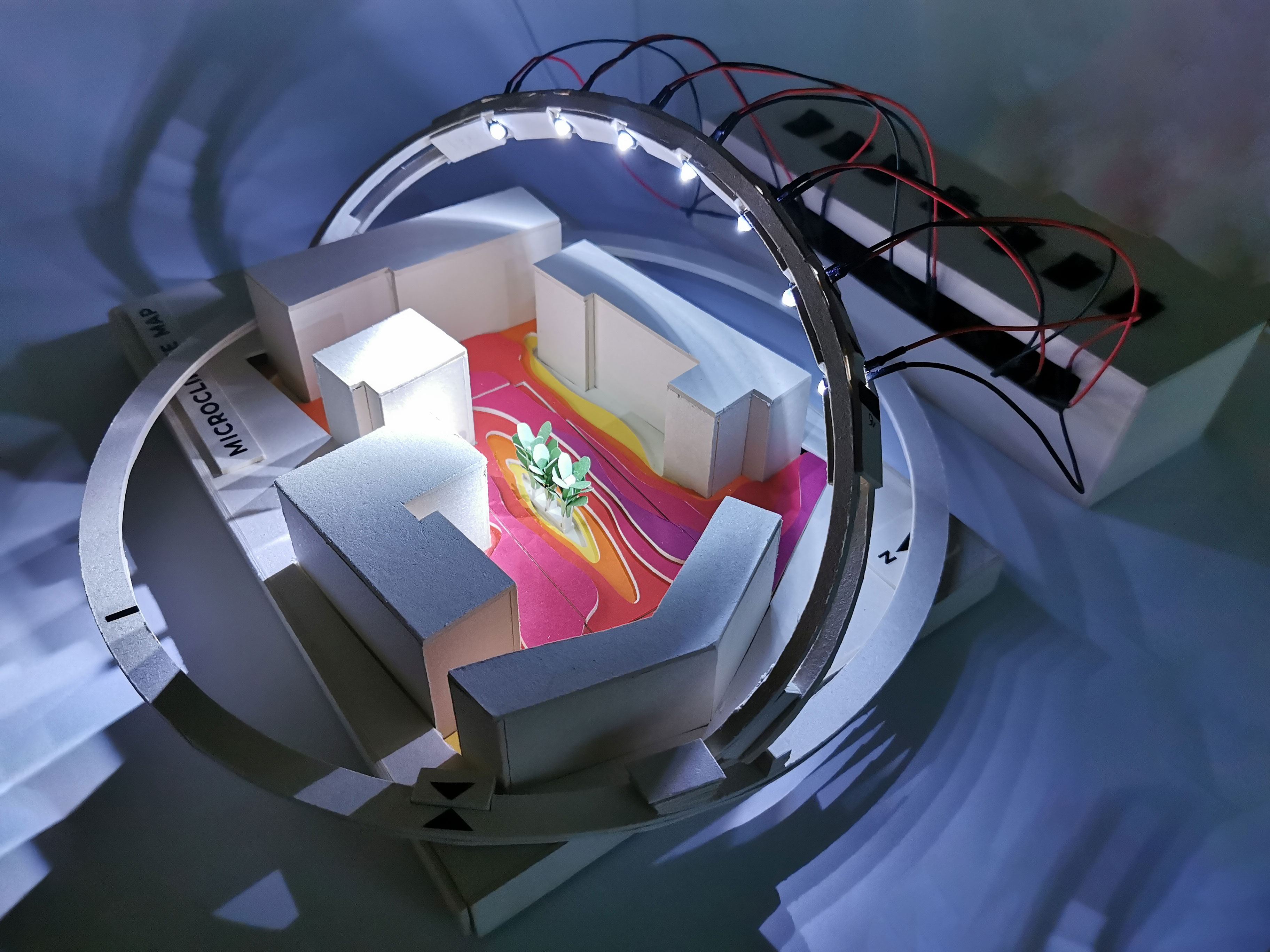}
    \end{subfigure}
 \caption{The physical model of the solar path in which the respective sun positions are represented by individual LEDs, each controlled by a dedicated switch: (top) off state; (bottom) on state.}
 \label{fig:solar_path}
\end{figure}

\paragraph{Context} Starting from the context that informs the visual design and functionality, our data physicalization encompasses critical fields of urban climate research, environmental communication, and situated data visualization. Since our primary audience is the general public, the final format should be familiar, intuitive, and easily relatable~\cite{schuster_designing_2026}. Hence, our physical model represents an urban neighborhood comprising built structures and open spaces, along with the respective tree models (Figure~\ref{fig:physicalModel}). In terms of functionality, the model should demonstrate the impact of local climate development driven by various climate adaptation strategies. We opted for widely adopted, cost-effective urban greening strategies. To further enhance the educational nature of our model, we have incorporated a tangible metaphor that represents a key factor in radiation exchange -- the sun's position along its path across the sky (Figure~\ref{fig:solar_path}).

\paragraph{Structure} The structure refers to the way the data is physically embodied and encoded, encompassing its material form and overall physical configuration. Here, we adopted the design decisions relating to fabrication from the architecture, specifically the construction of the physical terrain model. Hence, our physical model uses a layered contour model (i.e., a relief map) to encode the respective values, with temperatures encoded as heights of each layer (see Figure~\ref{fig:physicalModel}). Furthermore, the colors equally encode the respective values, creating a dual encoding system. To translate the digital data into their physical counterparts, we computed isolines from the numerical model output to identify the corresponding areas of equal value. The respective height layers are then positioned within their real-world context among the buildings, thus creating the final situated visualization. The solar path is represented as an external structure above the model, anchored at its natural angular position in relation to the ground surface on a given day of the year (Figure~\ref{fig:solar_path}). We used chipboard, a multi-layered recycled cardboard made entirely from waste paper, as the underlying material.

\paragraph{Interaction} The critical feature of our physicalization is the dialogue between the user and the model, which takes place through interaction at the artifact level. To achieve this, we considered two levels of interaction: direct interaction, involving the manipulation of the model's internal artefacts (see Figure~\ref{fig:userinteraction_model}), and indirect interaction, involving the manipulation of the external artifacts manifested through the solar path (see Figure~\ref{fig:userinteraction_LED}). For the former, we generated difference maps for each of the three scenarios and established spatial boundaries to distinguish areas of change from those with no variation. Users are invited to manipulate and exchange the respective boundaries of change, enabling them to better grasp and internalize the resulting impacts. For the latter, each sun position is represented by a single LED that can be switched on and off at will.

\subsection{Empirical study}

Informing the general public is a central goal of climate visualization~\cite{johannson2010}. However, the limited interactivity of commonly used methods~\cite{tominski_climate_2011} can negatively impact their readability and hide the complex nature of the data~\cite{Schuster2024}.
Data physicalization research has shown several key benefits over 2D visualization, including increased individual engagement~\cite{Pahr2024}, community engagement~\cite{Perovich2021a,Perovich2021b}, greater memorability~\cite{stusak_memorability_2015, Pahr2025b}, and improved spatial perception~\cite{jansen_evaluating_2013}. \textbf{Our aim was to investigate how physical embodiment and interaction shape user perception, behavior, and engagement in conjunction with urban climate data}. To this end, we conducted a controlled environment user study.

\paragraph{Metrics}

The main focus of our study was the qualitative analysis of user behavior. 
We used five \textbf{guiding questions} to elicit user feedback before and after the study (see Table~\ref{tab:questions}). We posed two pre-study questions to establish a baseline of participants’ existing knowledge regarding thermal conditions in cities (Q1) and the factors influencing them (Q2). After the study, we repeated Q1 and Q2 to see if participants gained insights or changed their opinions. Additionally, 3 questions were asked during the post-phase about the concrete measures presented (Q3), as well as experiences (Q4) and suggestions (Q5) on our presented physical model.

During interactions with the model, we encouraged participants to \textbf{think-aloud}.
To measure \textbf{subjective engagement} factors, we used the Affective Engagement Visualization Survey (AEVs) proposed by Hung and Parsons~\cite{Hung2018} and Hung~\cite{Hung2019}, a self-report instrument designed to quantify users' affective engagement during interaction with information visualizations (see Table~\ref{tab:AEVs}). The survey uses a 5-point Likert scale for evaluating each statement, with response options ranging from \textit{1 = Strongly disagree} to \textit{5 = Strongly agree}, allowing respondents to indicate varying degrees of agreement or disagreement.
Finally, we measured \textbf{interaction times} for our users at each stage of the experiment and \textbf{recorded the interactions} using a tripod-mounted camera (i.e., Zoom Q2n-4K video and audio recording device).

\paragraph{Recruiting}
Our recruitment process employed a convenience sampling approach, with participants drawn from readily accessible populations consisting of university students and staff members at our research institution. In the university context, participants were recruited from an ongoing Human–Computer Interaction (HCI) lecture taught by one of the authors. In addition, members of the research institution’s staff were invited to participate in order to broaden the diversity of perspectives and experiences represented in the study.

\paragraph{Procedure}

\begin{figure}[h!b]
    \centering
    \includegraphics[width=\linewidth]{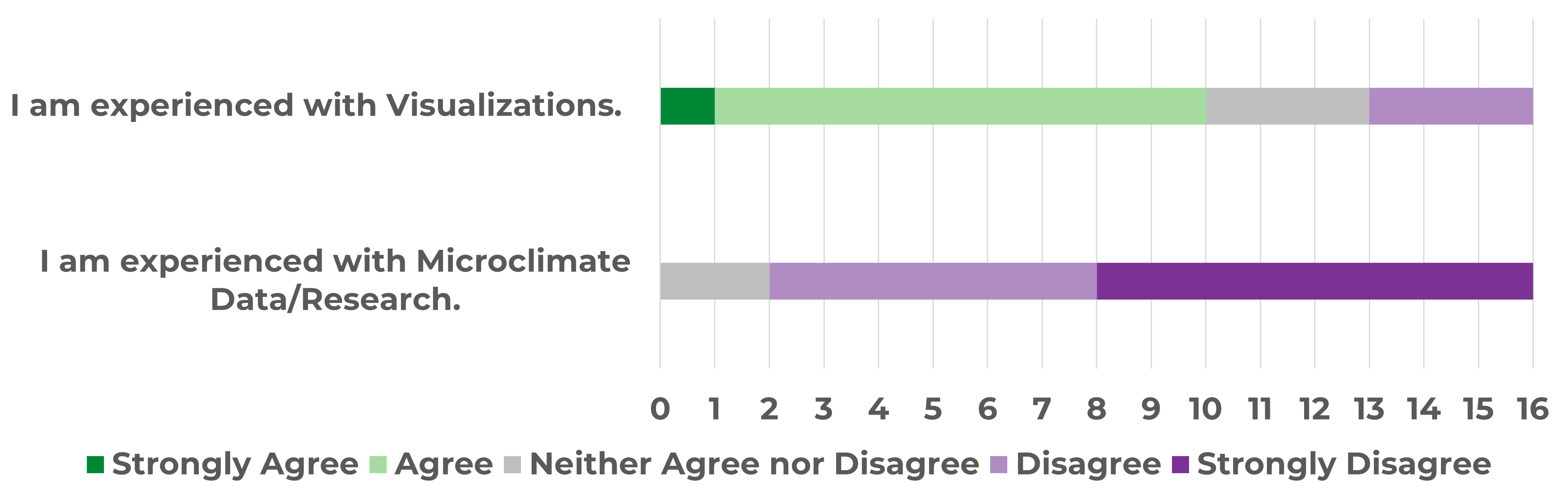}
    \caption{Overview of participants’ responses to the statements “I am experienced with microclimate data/research” and “I am experienced with visualizations,” measured on a 5-point Likert scale.}
    \label{fig:demographics}
\end{figure}

Study sessions started with us informing the participants of the purpose of the study. After obtaining written informed consent, we collected basic demographic information, along with background information on participants’ prior experience with microclimate data specifically and data visualization more broadly (see Figure \ref{fig:demographics}). We then proceeded with participant onboarding, during which we explained the study's focus and format. Specifically, we clarified that the study’s focus was on investigating different approaches to visualizing urban climate information in a selected urban neighborhood. Subsequently, participants were asked two pre-study questions (Q1 and Q2), as described in the Metrics subsection and presented in Table~\ref{tab:questions}. We then proceeded to the practical, hands-on component of the empirical study, employing a within-subject design in which the same participants were exposed to all experimental conditions (i.e., 2D map and the 3D relief model with and without LED lights).

Participants were presented with three different representations (i) 2D map, ii) 3D relief model, iii) 3D model with interactive lighting) depicting a selected climate adaptation scenario targeting heat mitigation in urban areas (i) without trees, ii) a single tree, and iii) three trees). 
Since our study was not focused on benchmark tasks, we decided to omit randomizing the order of conditions, in favor of gradually revealing additional features from $2D \to 3D \to \text{interactive 3D}$. This way, participants were introduced to the basic concept on the 2D variant and then to different levels of interactivity on the physical version.
 
\begin{figure*}[tb]
 \centering 
 \begin{subfigure}{0.32\textwidth}
        \centering
        \includegraphics[width=\linewidth]{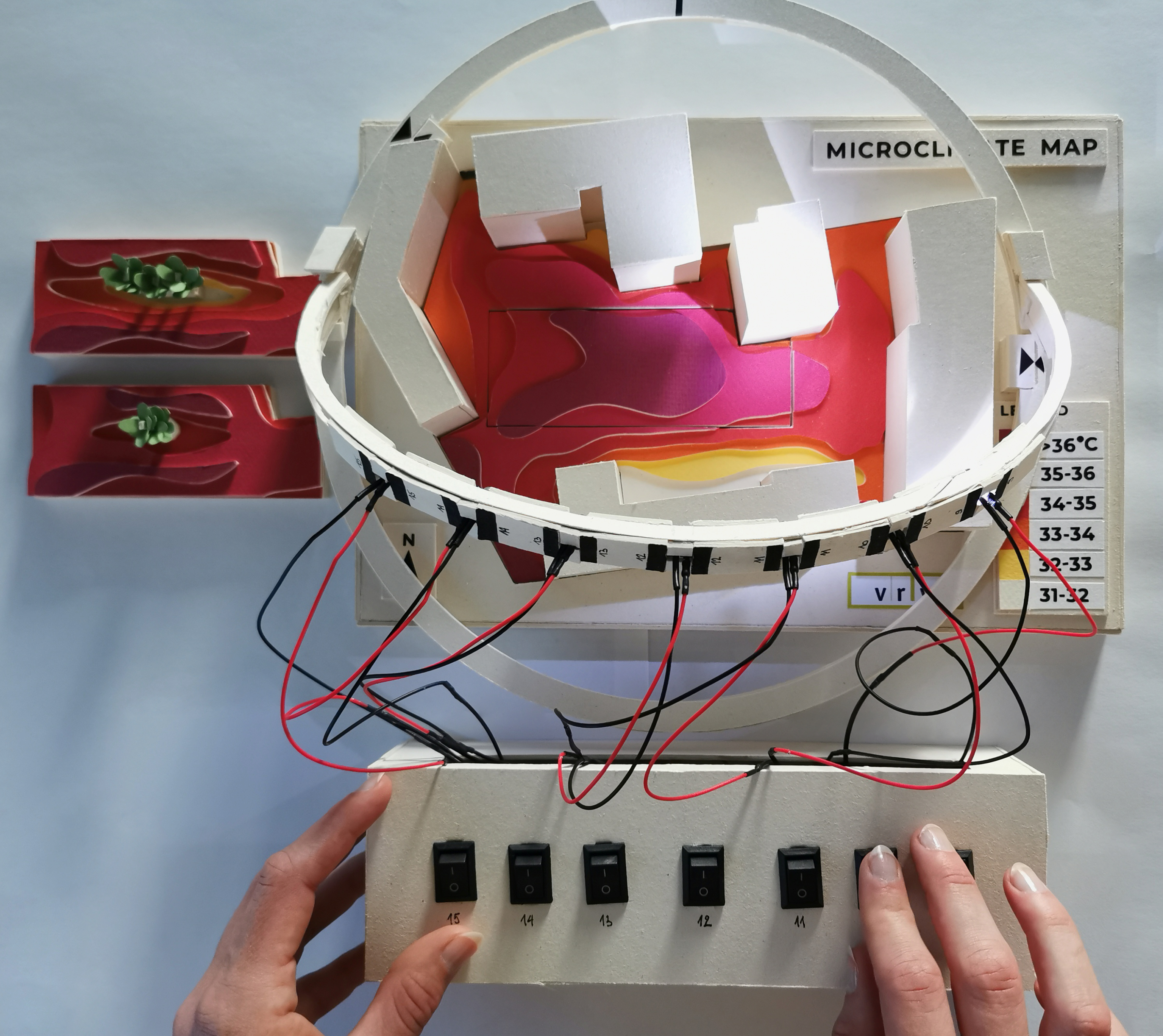}
    \end{subfigure}
    \hfill
\begin{subfigure}{0.32\textwidth}
        \centering
        \includegraphics[width=\linewidth]{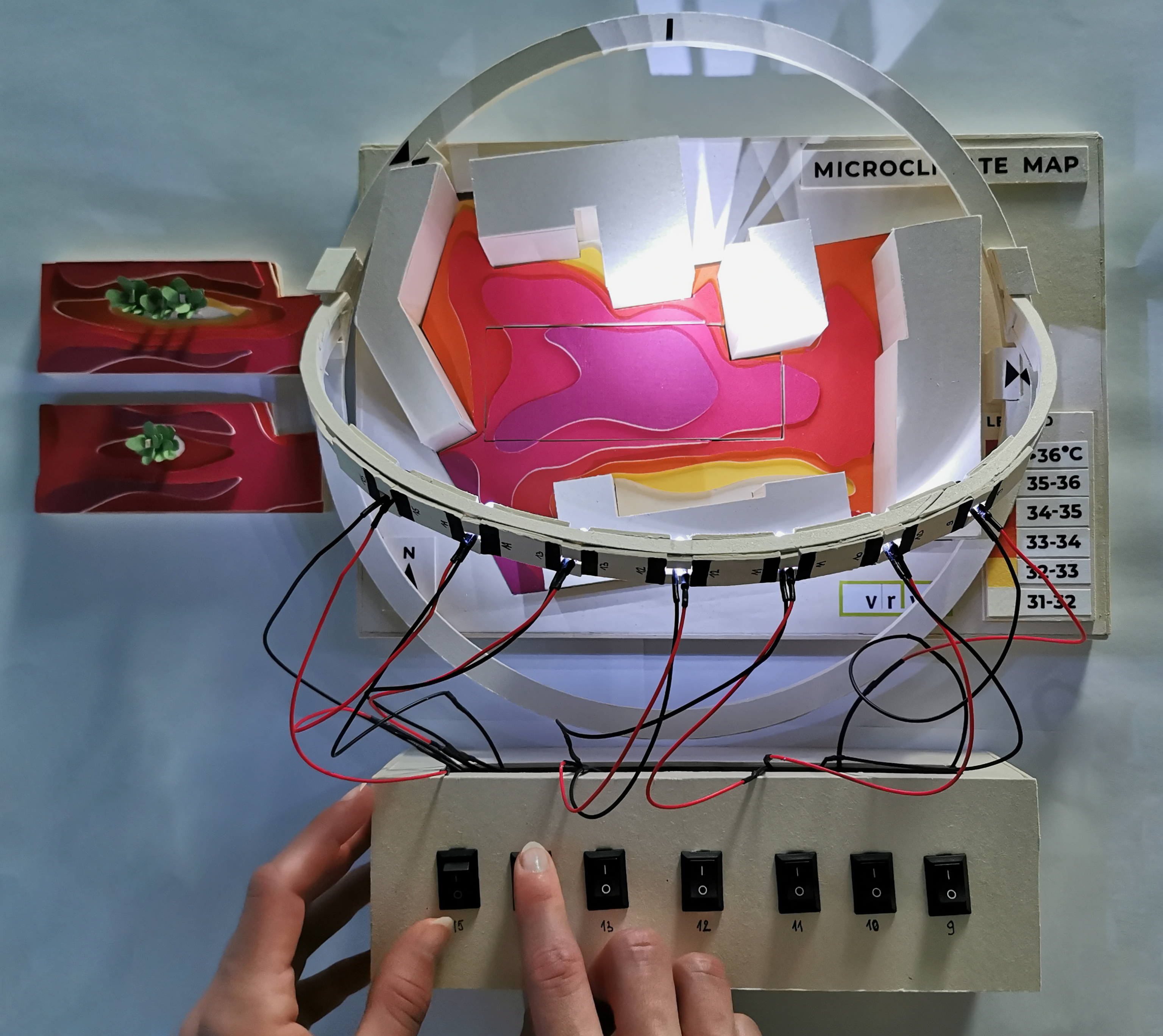}
    \end{subfigure}
    \hfill
\begin{subfigure}{0.32\textwidth}
        \centering
        \includegraphics[width=\linewidth]{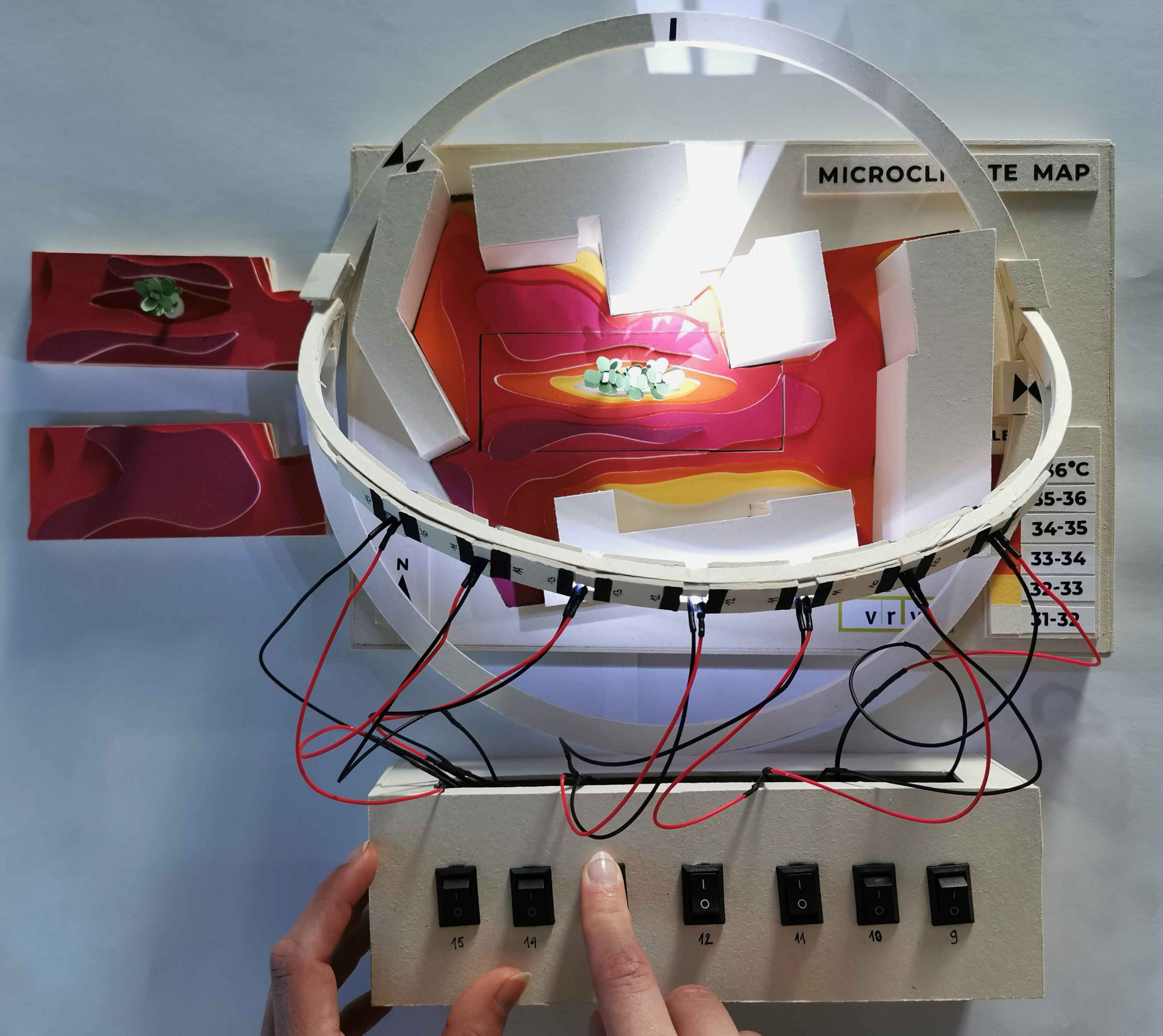}
    \end{subfigure}
 \caption{Sequential steps illustrating user interaction with the solar path model, showing how switches are used to control the corresponding sun positions by turning the LEDs on and off: (left) switching on the morning hours for the baseline condition; (middle) progressing to the afternoon hours for the baseline condition; (right) progressing to the afternoon hours for the multiple-tree condition.}
 \label{fig:userinteraction_LED}
\end{figure*}

During the initial exploration of the 2D map, The physical model remained covered and was revealed only after the 2D map task was completed. Participants subsequently interacted with the physical model. The interaction was structured into two distinct parts. 
The first sequence focused on direct interaction with the different exchangeable scenario inlays (Figure \ref{fig:userinteraction_model}), while the second introduced the solar path, prompting participants to engage with the model based on specific sun positions (Figure \ref{fig:userinteraction_LED}). 
Throughout the hands-on phase, participants were encouraged to think aloud and verbalize their actions, thoughts, and reasoning in real time as they explored the 2D maps and the physical model.

Following the exploratory hands-on phase, participants were presented with five post-study questions (see the Metrics subsection and Table~\ref{tab:questions}), two of which were repeated from the pre-study assessment to evaluate changes in reasoning. 
In the end, participants were asked to fill out the AEVs to rate their experience with the interactive physical model.

\begin{table}[tb]
  \caption{Overview of questions administered before (Q1-Q2) and after (Q1-Q5) the user study}
  \label{tab:questions}
  \scriptsize%
	\centering%
  \begin{tabularx}{\linewidth}{r X}
  \toprule
   no. & Questions \\
  \midrule
  Q1 & On a hot summer day, some streets and squares feel much hotter than others. What factors do you think influence spatial variation of heat across an urban neighborhood during a hot summer day? \\
  Q2 & What urban design strategies can help make outdoor spaces cooler and more comfortable for people? \\
  Q3 & How might trees, buildings, and open spaces influence temperature and shade in a city? \\
  Q4 & How was your experience with the model? \\
  Q5 & Do you have ideas on how we can further develop the model? \\
  \bottomrule
  \end{tabularx}%
\end{table}

\begin{table}[tb]
  \caption{The list of statements considered in the Affective Engagement Visualization Survey (AEVs).}
  \label{tab:AEVs}
  \scriptsize%
	\centering%
  \begin{tabularx}{\linewidth}{r X}
  \toprule
   no. & Item Statement \\
  \midrule
  1 & My use of this visualization is continuous and smooth. \\
  2 & I feel motivated while using this visualization. \\
  3 & I enjoy exploring this visualization. \\
  4 & I have acquired a new concept or new knowledge from the visualization. \\
  5 & I think the visualization effectively delivers its main concept or idea. \\
  6 & I think this visualization is telling a compelling story. \\
  7 & I think this visualization sparks my creative thinking. \\
  8 & I feel entertained when using this visualization. \\
  9 & I don't feel frustrated when using this visualization. \\
  10 & I feel absorbed by this visualization while using it. \\
  11 & The look and feel of the visualization is pleasing to me. \\
  \bottomrule
  \end{tabularx}%
\end{table}

\begin{figure*}[t]
     \centering
    \includegraphics[width=\linewidth]{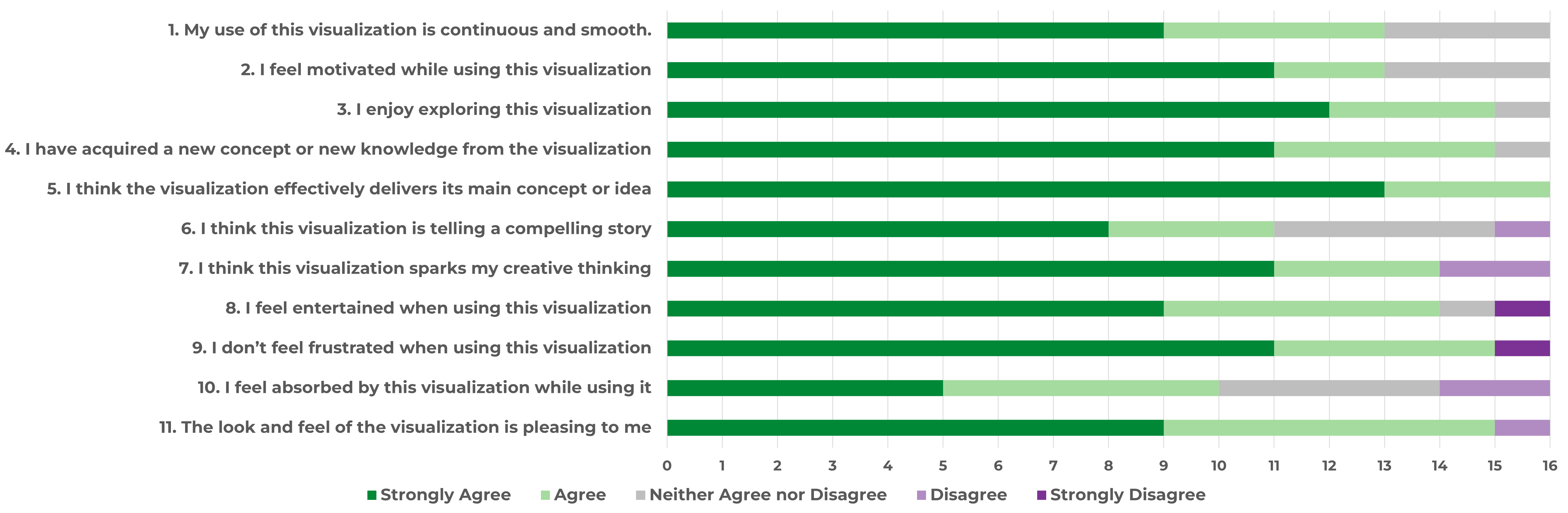}
    \caption{Overview of the participants’ responses to the Affective Engagement Visualization Survey, measured on a 5-point Likert scale.}
    \label{fig:AEVS}
\end{figure*}

\section{Results}

Given the small sample size (n=16), our analysis does not aim to produce statistically significant results. 
Instead, we focus on supporting qualitative data like interaction patterns and insights, using qualitative metrics, such as interaction times and subjective experience. With this, we seek to explore the potential applicability of a novel approach to visualizing climate data.

\subsection{Quantitative results}

In the context of the quantitative analysis, we aggregated responses from the 5-point Likert scale items in both the background questions (see Figure \ref{fig:demographics}) and the AEVs to assess participants’ \textbf{familiarity with the topic} and to measure \textbf{subjective engagement} factors, thereby facilitating comparison across questions and participants and identifying emerging tendencies within the dataset. We also recorded the time [mm:ss] each participant spent exploring each of the three different representations ($2D \to 3D \to \text{interactive 3D}$) to quantify \textbf{engagement duration}. 
Engagement duration was measured from the moment participants received the task instructions until they indicated that they were satisfied with their exploration.

\paragraph{Participants} 
We recruited a total of 16 participants (6 female, 10 male). As mentioned before, the majority of the participants (13) were university students who received course credit for an HCI lecture for participation. We recruited 3 additional female participants from our institution's research staff via convenience sampling. The participants ages ranged from 19 to 38 years (mean = 24.13, SD = 5.07). The highest level of education ranged from High School Diploma (12), Bachelor's Degree (1), Master's Degree (2), to Doctorate (1). 
With respect to \textbf{topic familiarity}, the group represented diverse backgrounds, with participants reporting some prior experience (n = 6) or a strong interest in data visualization (n = 10), and limited (n = 8) to no (n = 8) prior experience with microclimate data or research (see Figure~\ref{fig:demographics}).

\begin{figure*}[tb]
 \centering 
 \begin{subfigure}{0.32\textwidth}
        \centering
        \includegraphics[width=\linewidth]{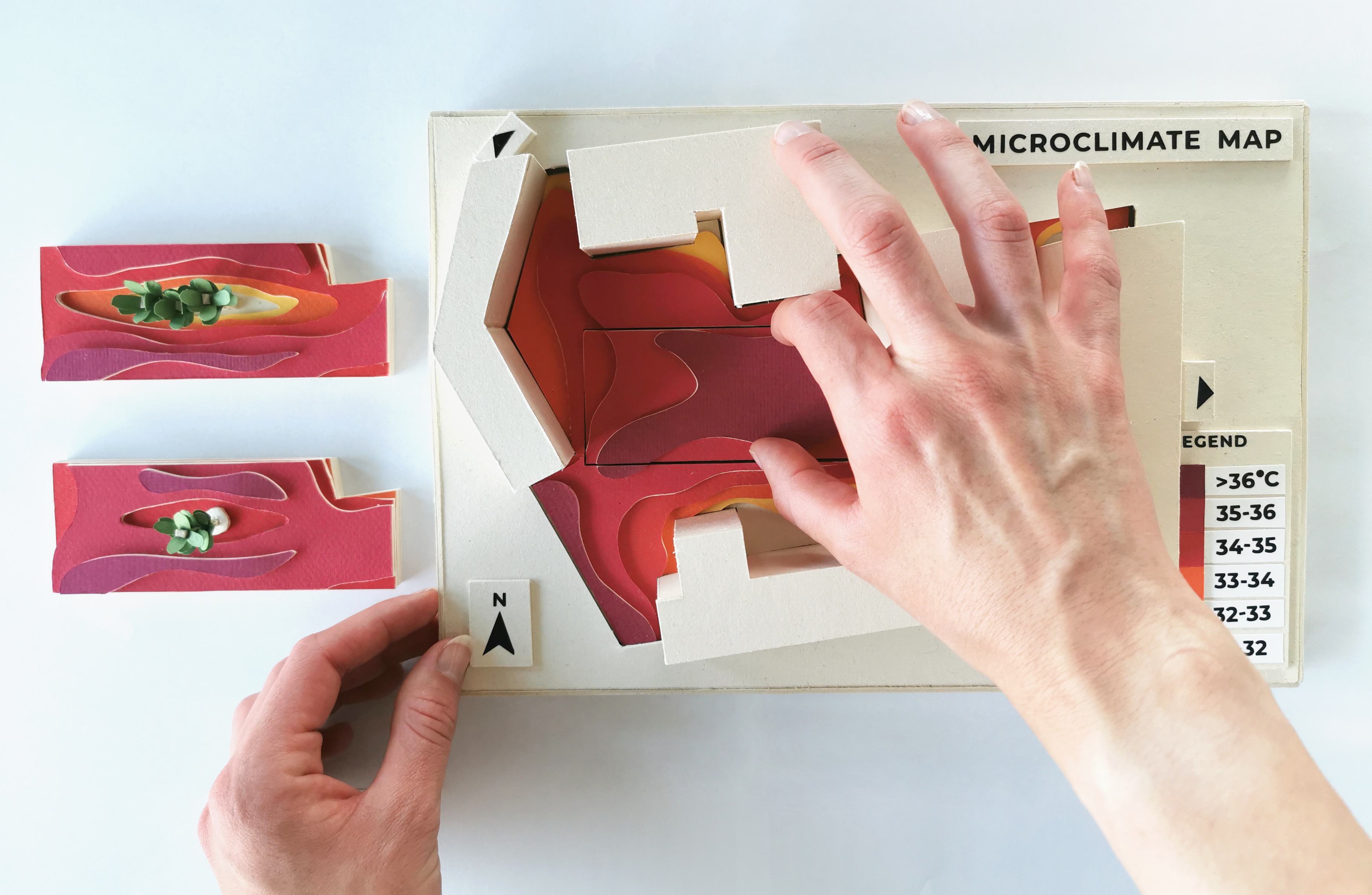}
    \end{subfigure}
    \hfill
\begin{subfigure}{0.32\textwidth}
        \centering
        \includegraphics[width=\linewidth]{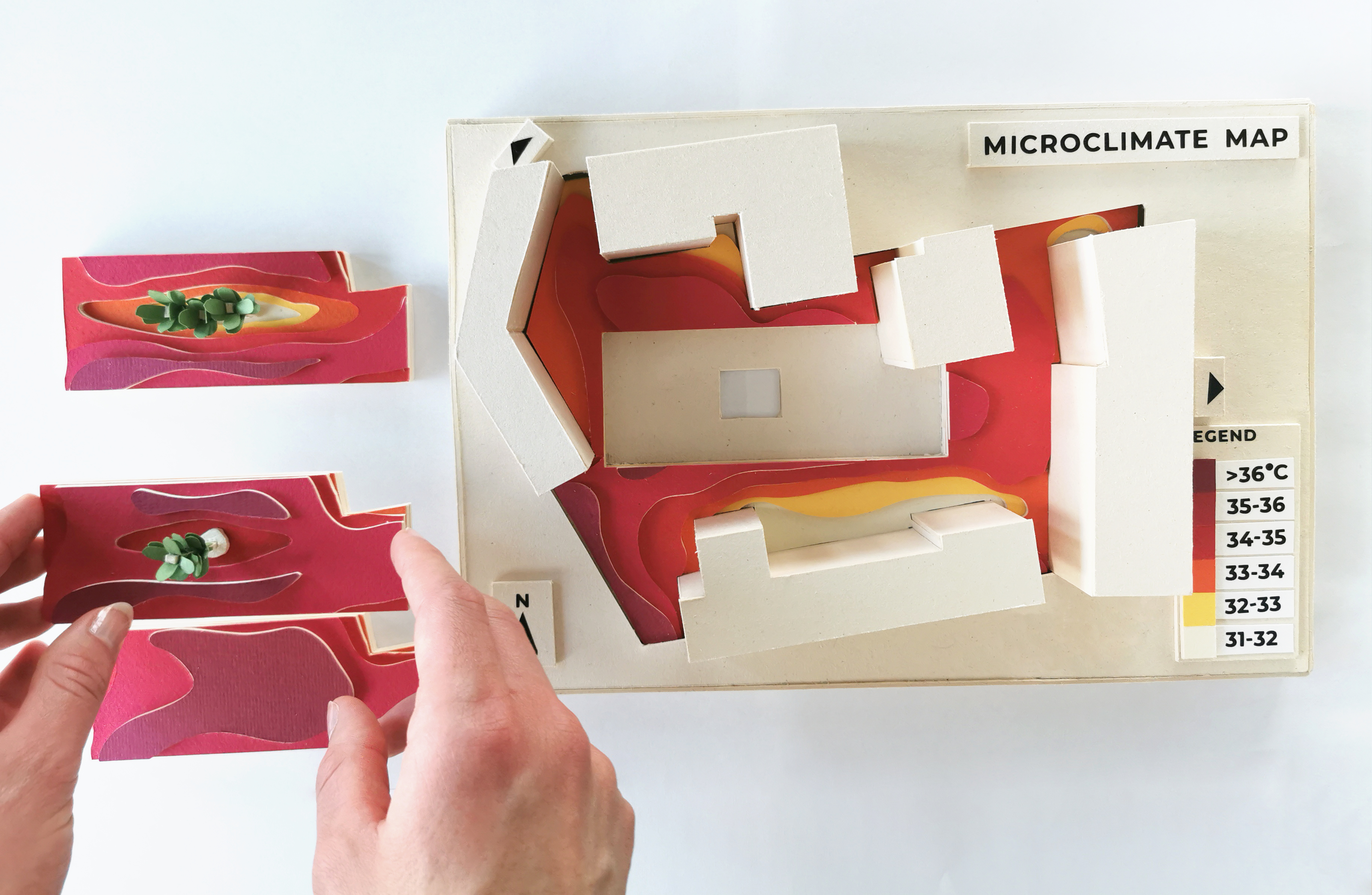}
    \end{subfigure}
    \hfill
\begin{subfigure}{0.32\textwidth}
        \centering
        \includegraphics[width=\linewidth]{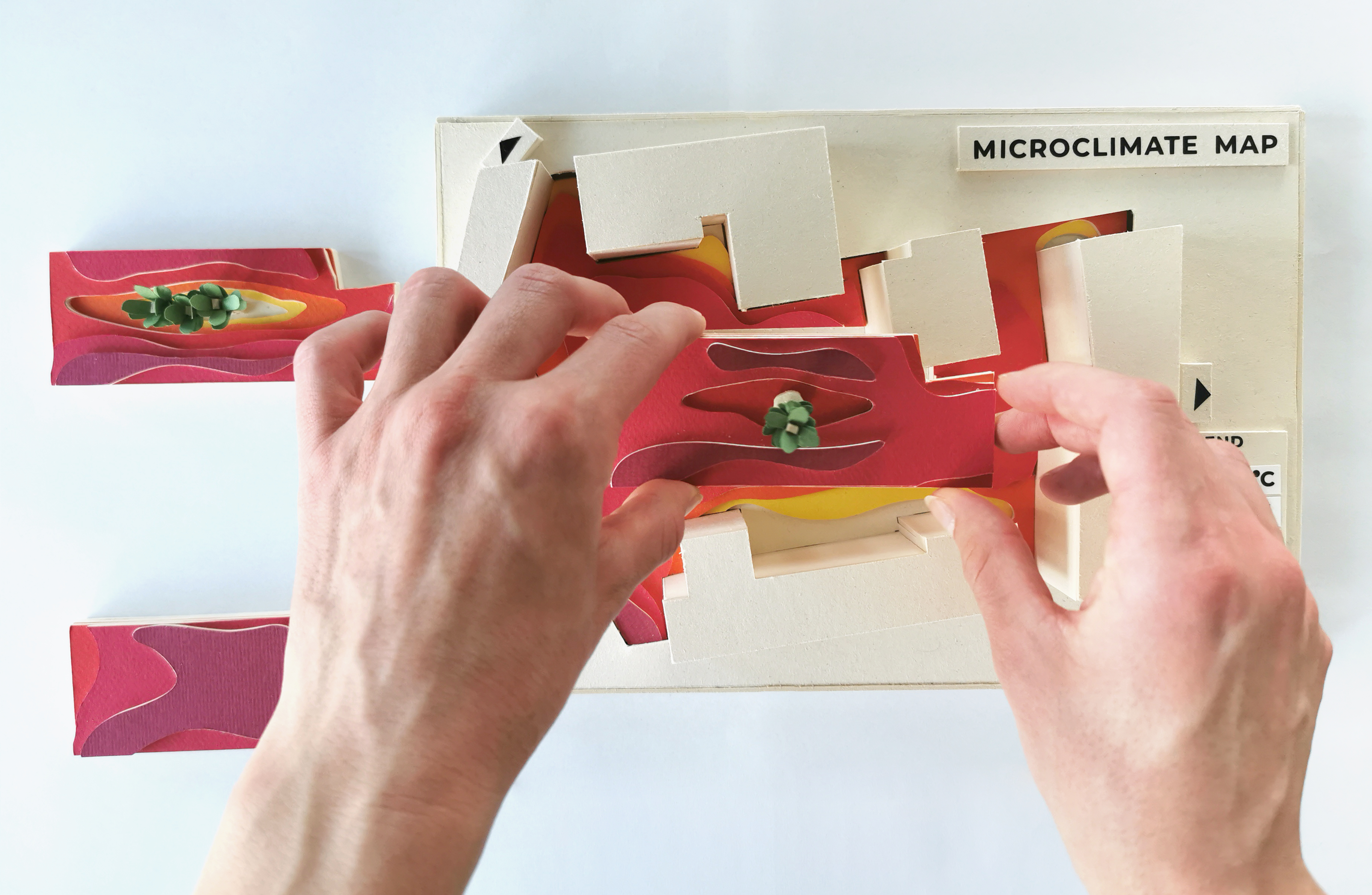}
    \end{subfigure}
 \caption{Sequential steps depicting user interaction with the physical model through the manipulation of exchangeable inlays representing different simulation outputs: (left) removing the baseline condition from the model; (middle) taking the single-tree condition; (right) placing the single-tree condition into the model.}
 \label{fig:userinteraction_model}
\end{figure*}

\paragraph{Engagement duration} 
On average, participants interacted with (i) the 2D map visualization for 02:37  (SD = 01:16), with (ii) the 3D relief model for 02:48  (SD = 01:02), and with (iii) the interactive lighting for 05:13  (SD = 01:51). 
However, several outliers were also observed. 
Five of the 16 participants spent less time interacting with the visualization (ii) than they did with the (i), one spent more time with (i) than (iii), and two spent more time with (ii) than (iii).

\paragraph{Subjective engagement}
An analysis of the AEV results indicates an overall positive reception of the presented visualization across the participant sample (Figure \ref{fig:AEVS}). 
All items except one were ranked strongly positive by at least half of the participants. All participants agreed that the representation \textbf{delivers its concept effectively} (Item 5). Additionally, we found strong agreement that participants \textbf{enjoyed} (Item 3) exploring the physicalization and \textbf{learned a new concept} or gained new knowledge through it (Item 4).
While the median of participants still agreed that they \textbf{felt absorbed} (Item 10) by the representation, this item was consistently rated lower than the others. Finally, we acknowledge that items referring to the representation's ability to \textbf{promote creative thinking} (Item 7) and creating \textbf{a feeling of being absorbed} received negative ratings from two participants.

\subsection{Qualitative results}

We report the average interaction time per condition, along with the standard deviation in [mm:ss].
In the context of the qualitative analysis, we analyzed participants’ open-ended responses (questions Q1-Q5) and drew on video recordings and field notes to identify \textbf{recurring themes} emerging during their interaction. 
We further analyzed the video recordings to categorize participants’ \textbf{interaction behaviors} during the tasks. This involved systematically reviewing the video footage to identify and code distinct types of engagement, such as exploratory actions, comparative viewing strategies, instances of passive observation, and moments of verbalized (aloud) reflection.
We report the common themes in \textbf{bold} text, as well as their relative occurrences in brackets (n), counting each theme only once per participant, regardless of how many times it was mentioned by the same participant. Example quotes paraphrased for readability are provided in \textit{cursive}.

\paragraph{Urban Climate Awareness (Q1-Q3)}

\begin{figure}[tb]
 \centering 
 \begin{subfigure}{0.45\textwidth}
        \centering
        \includegraphics[width=\linewidth]{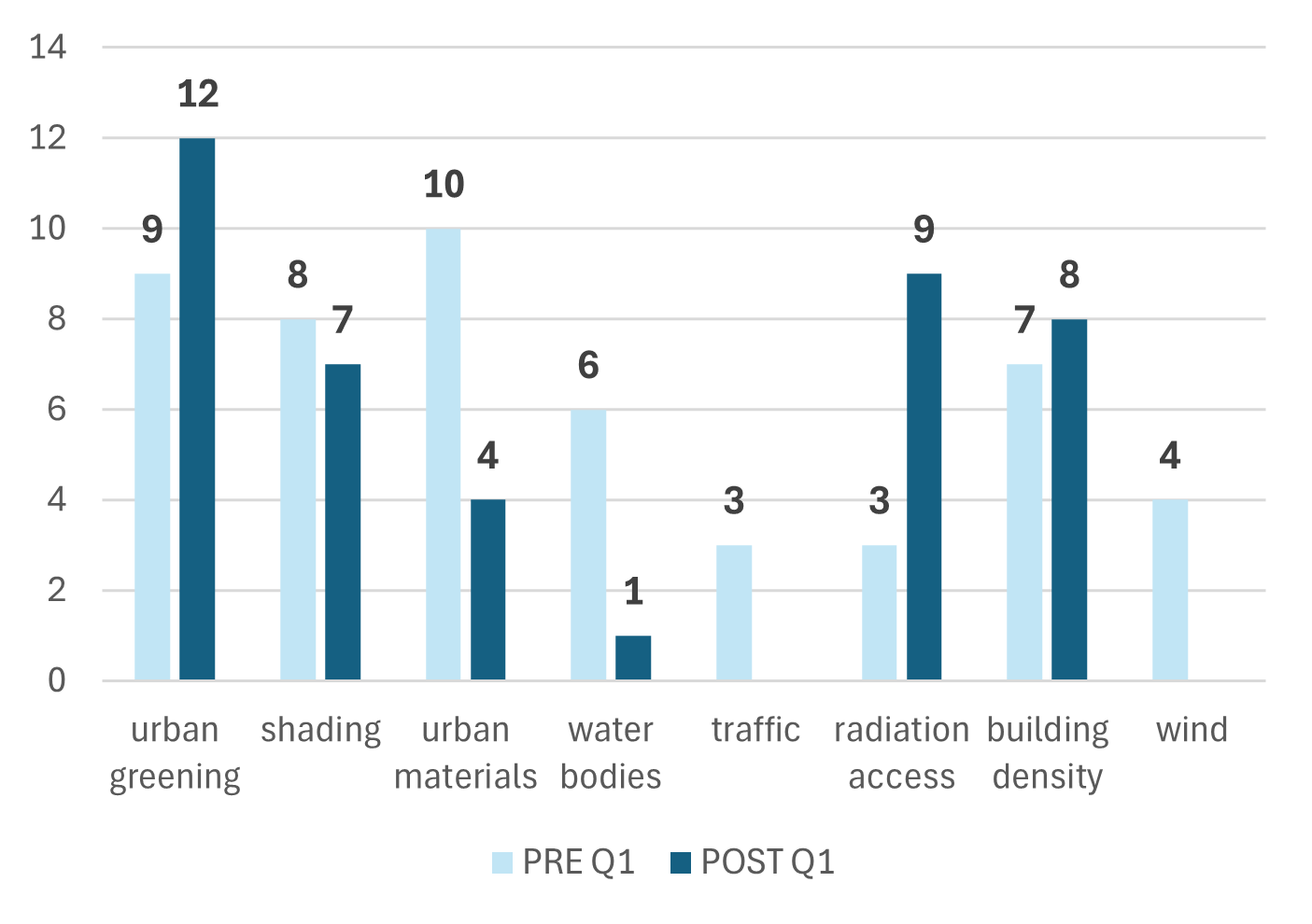}
    \end{subfigure}
    \vfill
\begin{subfigure}{0.45\textwidth}
        \centering
        \includegraphics[width=\linewidth]{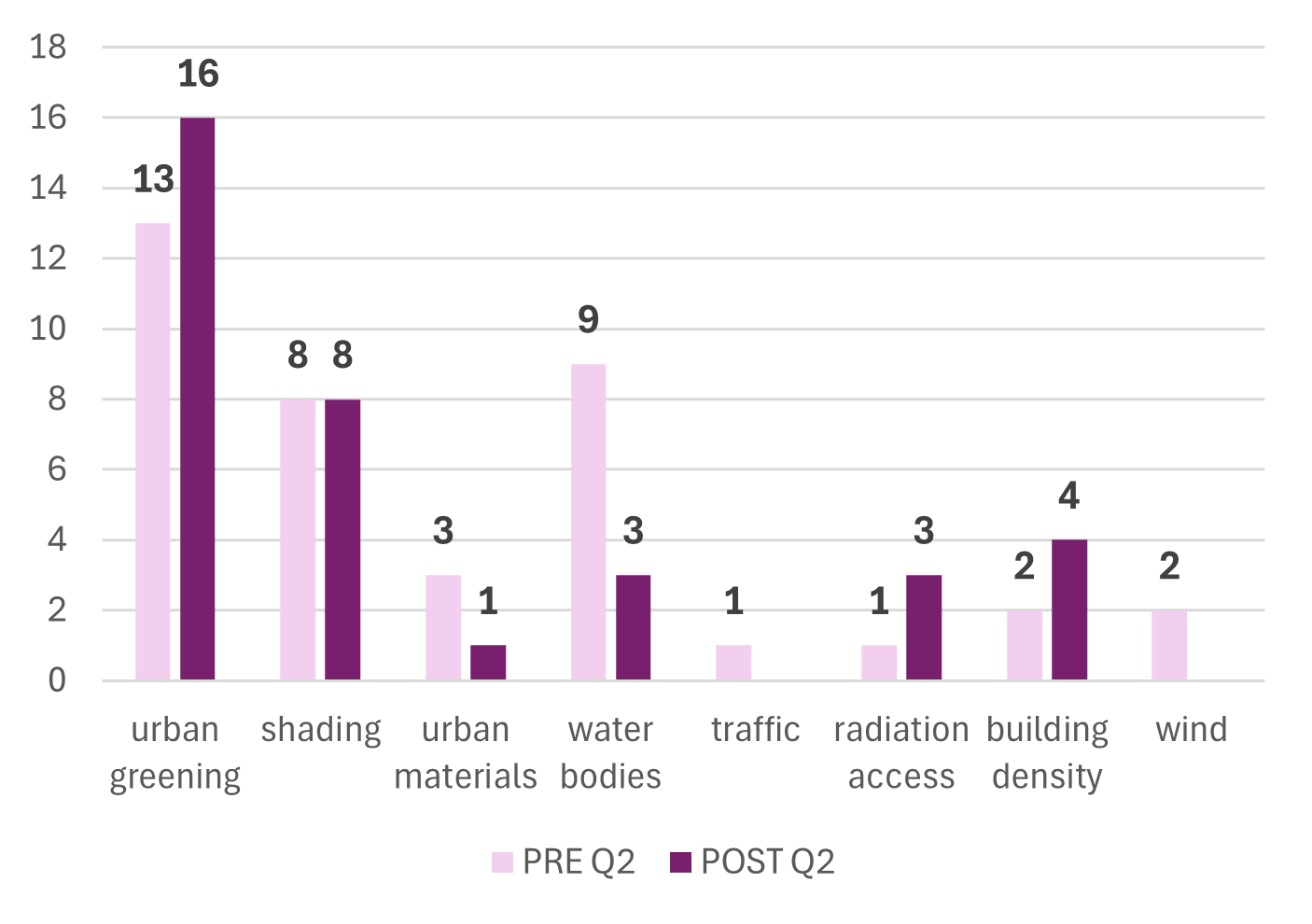}
    \end{subfigure}
 \caption{A cumulative overview of participants’ responses to questions Q1 (above) and Q2 (below) before and after the exploration period, respectively, classified by the identified recurring themes.}
 \label{fig:Q1_Q2}
\end{figure}

Looking at the pre-study phase, responses to Q1 show a strong prevalence of concepts related to \textbf{urban greening} (9), \textbf{shading} (8), and \textbf{urban materials} (10). A similar pattern emerges in Q2, where \textbf{urban greening} (13), \textbf{shading} (8), and \textbf{water bodies} (9) are prominently represented. These recurring themes indicate a strong collective awareness of how environmental and material factors contribute to the overall thermal situation of urban spaces. In the post-study phase, participants demonstrated a clear shift in perceptions, with greater emphasis placed on \textbf{urban greenery} (12), \textbf{access to solar radiation} (9), and \textbf{building density} (8) in the context of Q1. The same concepts are also prominently reflected in responses to Q2 in the post-study phase. These patterns indicate the integration of additional considerations into participants’ conceptual and sensemaking frameworks, primarily driven by insights from the physical model exploration, which were clearly reflected in several participants’ responses. We show how participants' answers changed between pre- and post-interaction interviews in Figure~\ref{fig:Q1_Q2}.
For instance, our records include the following paraphrased account of one participant’s statement, corresponding to the concept of \textbf{shading}: \textit{“The sun’s position is now more apparent, something I hadn’t considered before, and you can see the different effects throughout the day.”} Additionally, we recorded the following paraphrased statement regarding heat distribution, corresponding to the concept of \textbf{access to solar radiation}: \textit{“I would expand on my previous answer, as I now observed that some of my assumptions were correct, namely, that large open spaces tend to generate more heat, and without any mitigating elements, heat can easily build up in these areas.”} 
While 4 participants explicitly stated that they had already covered everything in their responses to the previous questions and therefore did not answer Q3, the remaining 12 reinforced their belief in the beneficial effects of greenery, building shade, and the configuration of open spaces that can facilitate airflow.

\paragraph{Participant Experience (Q4)}
In response to Q4, we received predominantly positive feedback on the physical model from participants, with 15 clearly stating that they liked the model very much and perceived clear benefits from this form of representation. To exemplify, we are including a paraphrased statement: \textit{“I liked the physical model a lot because I understand the data better when I can touch and interact with it, rather than just viewing a printed graphic."} Another participant stated: \textit{"I really like the model. I saw this kind of model for the first time, and I like that it is hand-crafted."}. Only one participant remained unconvinced by the third dimension, questioning the added value compared to the effort required to build such a model versus using its digital counterpart or a simple 2D map. Nevertheless, the remaining participants provided positive feedback on the visualization, particularly noting its 3D dimension (7), encoding using isolines (3), the possibility of exchanging scenarios (5), and the ability to manipulate light and shadows using the LED setup (9).

\paragraph{Improvement Suggestions (Q5)}
In response to Q5, recurring themes were once again identified across participant responses. These themes primarily relate to the \textbf{consideration of alternative scenarios} (10) and \textbf{adjustments to the solar path model} (7). The former includes strategies such as increasing vegetation (e.g., planting additional trees), modifying building layouts, geometries, and facades (e.g., adding windows), and incorporating shading elements such as canopies. The latter involves enhancing the adaptability of the solar path representation, for instance, by adjusting its angle seasonally and enabling lighting elements (i.e., LEDs) to vary in intensity according to the time of day. A less frequently occurring theme concerned the incorporation of \textbf{heat simulation} (2) to enhance the model’s tactile qualities and improve accessibility, particularly for individuals with sensory sensitivities, as suggested by two participants.

\paragraph{Interactions With the 2D Map}
We observed two distinct approaches that participants used to explore the 2D map: \textbf{sequential progression} from the baseline through the various tree configurations (11), and the \textbf{simultaneous examination} of all three maps to facilitate instant comparative assessment (5). Within the first approach (sequential progression), additional behavior patterns were identified: four participants advanced to subsequent scenarios without revisiting prior ones, whereas seven repeatedly returned to earlier scenarios to enable comparative analysis. Regarding insight generation and sensemaking, all 16 participants engaged with the baseline conditions on the 2D map mostly through color encoding, relying on abstract interpretations, such as interpreting color as a proxy for intensity, for example, associating stronger red hues with heightened (i.e., hot) conditions: \textit{"The very dark colors in the middle mean it is very hot"}. Eight of the 16 participants also consulted the numeric temperature values provided in the legend, particularly when examining scenarios with single- or multiple-tree configurations. Nine participants also made observations regarding the buildings and their shading effects, three of whom questioned whether the white blocks represented buildings at all (see Figure \ref{fig:sims}).

\paragraph{Interactions With the 3D Model}
Looking at interactions with the physical model, the majority of participants (12) followed the intended interaction sequence, progressing from the baseline condition through the different scenarios by directly replacing the exchangeable inlays in the physical model, moving from \textbf{single-tree to multiple-tree} configurations (Figure \ref{fig:userinteraction_model}). Three participants \textbf{skipped the intermediate} step and moved directly from the baseline to the multiple-tree configuration, while one participant examined the inlays only outside the main model by placing them side by side. In terms of verbal feedback, eight participants reacted immediately upon uncovering the physical model and explicitly praised its appearance, describing it as a clearer and more visually appealing representation of the conveyed information: \textit{"I like this model a lot!"}, \textit{"It is much more haptic."}. After replacing the baseline condition with tree configurations, 14 participants explicitly stated that the thermal effect was more visible in the isoline representation in both single- and multiple-tree scenarios than in the 2D map, while 7 participants also noted a relatively localized, less pronounced effect of a single tree compared to multiple trees. One outlier was observed, with a single participant indicating that the model provided no additional information beyond what the 2D map already conveyed. 

\paragraph{Interactions With the LED Model}
In terms of LED interaction, two distinct behavioral patterns were observed: a \textbf{sequential progression} across hours, in which the LED corresponding to the previous hour was turned off (11), and an \textbf{accumulative pattern}, in which LEDs were activated one by one while all previously activated LEDs remained on (4). However, three deviations in initial interaction were also observed: (a) activating only the LED corresponding to 12 PM (i.e., midday/noon), (b) activating LEDs at the temporal extremes (i.e., earliest and latest hours), and (c) randomly switching LEDs on. Following these initial behaviors, these three participants subsequently converged on one of the primary interaction patterns described above. In sequential exploration, one participant indicated that this interaction facilitated understanding of the effects of individual sun positions, stating that it allowed them \textit{“to see how the sun moves”}. In contrast, another participant, who had all lights switched on at once, reported that the interaction supported comprehension of cumulative sun exposure, noting that \textit{“if you switch on all suns, you get a kind of a heatmap where brighter spots are exposed longer}. Following the initial LED interaction, 14 participants explored the effects of different inlays by either exchanging all three inlays or testing a single alternative. Subsequently, they examined individual LED positions according to one of the two primary interaction patterns. Only two participants did not engage with additional inlays beyond the one initially present in the model. We also observed a preference for exploring the impact of sun position, with 10 participants focusing on the radiation effect around noon and in the late afternoon. As one participant noted, this gave them a better understanding of the sun's full exposure driven by \textit{“the sun's angle and position on the sky”}. Another participant observed that only after 11 AM \textit{“the impact of tree shadow becomes visible and from 12 PM we see the direct impact on trees”}.

\subsection{Discussion}
We conclude our analysis by translating our qualitative and quantitative results and discussing how our situated geospatial thermal physicalization prototype can inform future work.

\paragraph{Insights Through Hands-on Exploration}
While participants were able to reason about the presented 2D maps competently, we find that physical interaction with our model facilitates not only user engagement but also insight generation, consistent with past results~\cite{Pahr2024, hurtienne_move_2020}. 
Our participants followed a variation of interaction patterns, motivated by the physical affordances of our prototype, and expressed interest in increasing the model's complexity to demonstrate different scenarios on a larger scale. This highlights the potential of data physicalizations as public displays of information that support self-directed exploration.

\paragraph{Multifaceted Engagement}

Climate visualization is an important tool to raise public awareness~\cite{johannson2010}, an aspect usually disregarded in performance and efficiency-based evaluations~\cite{wang_emotional_2019}. Our AEVs~\cite{Hung2019} results show that our prototype was well received regarding most engagement aspects, highlighting its potential to support conceptual understanding through effective, engaging, and visually appealing storytelling.
While immersion into the prototype was a notable outlier and not rated as positively as other engagement aspects, we also observed longer interaction times with our physical models, indicating that a greater number of meaningfully interactive features can motivate observers.
Verbal feedback during interactions indicated that engagement was supported by both physical interaction and the novel visualization concept.
Despite the overall positive results, we note that one participant was generally unimpressed by the representation and conclude that no one-size-fits-all solution exists for environmental public communication.

\paragraph{Shifting of Perspectives}
We observed general improvement in reasoning about the factors influencing heat distribution across the three leading questions in the post-study phase. 
Participants generally demonstrated a deeper understanding of the underlying mechanisms, particularly regarding the role of solar exposure and related urban heat patterns, as well as the importance of tree placement, building orientation, and the openness of the space. 
Specifically, they were able to articulate why certain areas within the study domain consistently remain hot, while others remain shaded. 
In the past, physicalizations have been used to raise awareness in specific communities through situated displays~\cite{Perovich2021a} and workshops~\cite{Perovich2021b}.
While our physicalization prototype consisted of a single physical model and our study evaluated user interactions in an individual setting, our approach can be easily employed in group or public settings to raise awareness on urban climate adaptation strategies.

\section{Critical Reflection, Limitations and Future Work}

\paragraph{Participant sample} Our study population was recruited from computer science students and visualization experts. However, most studies draw on a similar sample even when their focus is on the general public~\cite{Burns2023}. Yet, this poses a key limitation: our participants were mostly young, educated, and predominantly male. 
Nevertheless, the gender balance was mitigated through targeted recruitment of female participants via convenience sampling. However, all participants recruited in this way were still highly educated peers.

While our study constitutes a first step in investigating the effects and potential benefits of physicalizing thermal data for public communication, we acknowledge that a more diverse sample is needed to make broader claims.
Older or younger individuals, especially given their different technological expertise or education levels, may interact differently with the proposed physicalization model, and thus these aspects require further investigation.
Especially community-relevant topics, such as urban microclimate heat mitigation, can profit from workshop-based formats involving local residents of the sampled neighborhood~\cite{Perovich2021a,Perovich2021b}. Such formats would bring people directly affected by proposed measures into contact with the data, possibly sparking conversation and increasing awareness of the topic.

\paragraph{Scalability and transferability} For the creation of our three climate adaptation scenarios, we only considered a small, locally constrained area of the city. This is also reflected by the function and placement of the ``halo'' in the interactive physical model. While the local simulation remains reasonably accurate, it ignores buildings and vegetation outside of the presented area. 
For simulations of larger urban areas, the localized exchangeable inlays with relief maps would be of limited use, and a larger ``halo'' to physically represent lighting conditions would require greater technical finesse. 
To still make use of the positive effects of physical interactions, the shadow maps could be created by projecting UV light on photosensitive, color-changing material, similar to the method proposed by Patnaik et al.~\cite{patnaik_sensemaking_2023}. This would allow users to investigate the impact of heat-mitigating techniques on urban areas in real time.

Applying the same research design to a larger area would raise two considerations. First, alternative fabrication technologies such as 3D printing might be needed to ease the model-building process. Second, for more complex simulations, such as representing daytime and nighttime conditions, we would choose to design the entire thermal map as a removable component, allowing different maps to be taken out and replaced by a new condition.

\paragraph{Model adaptability} While our physical model dynamically simulates the movement of incoming radiation and the resulting shadows throughout the day via the LED apparatus, the physical temperature layers remain static and do not respond to the changing sun position. In contrast, a purely digital model could continuously update the simulated temperature distribution to reflect each temporal step. This decoupling between dynamic insolation and static thermal visualization constitutes one of the key methodological differences from fully digital approaches. Future iterations of the physical model could mitigate this limitation, for example by incorporating thermochromic materials or a hybrid setup in which the physical model interfaces with a digital twin that recalculates temperature layers in real time.

\paragraph{Evaluation metrics} Finally, our study focused heavily on qualitative aspects to deliver valuable insights with a relatively small sample.
Our study design did not involve structured benchmark tasks, and we omitted randomization of experiment conditions.
We argue that, for a technique suited to public engagement through interactive data displays, free exploration tasks are well-suited to demonstrate their intuitivity and potential~\cite{north_insight_2006}.
It has been argued that efficiency-based metrics may not capture the full benefits of physical data representations~\cite{wang_emotional_2019}.
Still, it may be valid to investigate task-based efficiency metrics, such as accuracy and task completion times. Additionally, value-based subjective metrics, such as ICE-T~\cite{wall_icet_2018}, could provide further insights into how physical representations compare to virtual counterparts.

\section{Conclusion}

We introduced a novel approach to climate communication and education through data physicalization and hands-on engagement. Specifically, we provided users with an situated geospatial thermal physicalization prototype that integrates selected climate adaptation strategies (i.e., tree planting) as exchangeable inlays, alongside an externally mounted, electronically driven solar path with interactable suns, represented by multiple discrete LEDs. Through direct, hands-on interaction, we investigated users’ responses to and engagement with the physicalization prototype, observing a high overall proportion of favorable reactions.
With some modifications, our interactive microclimate model can be adapted to investigate larger areas and facilitate community engagement, further highlighting the potential of data physicalizations for climate visualization.

\acknowledgments{
The VRVis GmbH is funded by BMIMI, BMWET, Tyrol, Vorarlberg, and Vienna Business Agency in the scope of COMET - Competence Centers for Excellent Technologies (904918, 911654), which is managed by FFG. This work has additionally been funded by the Vienna Science and Technology Fund (WWTF)
[Grant ID: 10.47379/ICT25043].}

\bibliographystyle{abbrv-doi}

\bibliography{template}
\end{document}